\documentstyle[aps,preprint,psfig]{revtex}
\tightenlines
\begin{document}
\draft
%\twocolumn
\title{Measurement and state preparation via ion trap quantum
computing.}
\author{S.~Schneider, H.~M.~Wiseman, W.~J.~Munro, and G.~J.~Milburn}
\address{Department of Physics, The University of Queensland, 
St. Lucia, Brisbane, Qld 4072,
Australia}
\maketitle
\begin{abstract}
We investigate in detail the effects of a QND vibrational number
measurement
made on single 
ions in a recently proposed measurement scheme for the vibrational 
state of a register of ions in a linear rf trap [C.~D'Helon and
G.~J~Milburn, 
Phys.\ Rev.\ A {\bf 54}, 5141 (1996)]. The performance of a measurement 
shows some interesting patterns which are closely related to 
searching. 
\end{abstract}
\date{today}

%%%%
%%%%%%%%%%%%%%%%%%%%%%%%%%%%%%%%%%%%%%%%%%%%%%%%%%%%%%%%%%%%%%%%%%%%%%%%%%

\narrowtext

\section{Introduction}

The realisation that quantum physics enables a computer to solve certain
problems more efficiently
than a classical computer\cite{ekert} has led to a revision of our
understanding of information
processing in physical systems. The key feature of quantum physics which
lies behind the new
developments is, as usual, the superposition principle. In a quantum
computer certain properties 
of a function can be determined, in a single step,  by evaluating the
function on a coherent
superposition of input states. A classical computer in contrast would
need
to evaluate the function
on separate inputs one step at a time. While we are a long way from a
complete
understanding and classification of quantum algorithms, a number of
general
features may be
distinguished. Firstly, some part of the physical computer, the input
register,  is used to encode
the input states. Typically this involves a set of $N$ two-state systems
which may be each prepared
in an arbitrary superposition state. Each of these systems is said to
encode
a qubit, as distinct
from the single bit encoded in a classical two state system. As the
product
Hilbert space of the
input system is of dimension $2^N$, it may be used to encode a massive
amount of information in a
single large superposition state. Secondly we identify another part of
the
physical computer, the
output register,  to encode the output states. The output register may
likewise be a set of
two-state systems. Finally there is some unitary transformation which
correlates the states of the
output register and the input register in such a way that the output
register now encodes the
result of some function evaluation of the input. The final result is a
large
entangled state of the
input and the output registers. This unitary transformation is a very
high
dimensional object, and  
would in general require a very complicated time dependent Hamiltonian.
However we now know that an
arbitrary unitary transformation may be approximated by a suitable
sequence
of simpler
transformations involving at least two qubits \cite{ekert}. 

The scheme described in the previous paragraph has a lot in common with
the
von Neumann model of a
quantum measurement. Instead of the input register, we speak of the 
{\em
measured system}, and
instead of the output register we speak of the {\em apparatus}. However
in
the von Neumann scheme
traditionally, we have only considered unitary transformations generated
by
a single time
independent  Hamiltonian. Having seen the quantum computation scheme,
this
seems to be unnecessarily
restrictive. We thus have a novel view of a measurement  as an
information
processing system in
which information in the measured system is processed and written to the
apparatus. The
possibility of choosing a much larger class of unitary transformation so
to
couple the system and
apparatus, and to further process information in the apparatus before
read-out, is the motivation of
our work.

In a recent paper \cite{dhelon} a method was proposed to make an
approximate quantum nondemolition measurement of the number state
distribution of
the  collective vibrational mode of a $N$-qubit ion register. In that
paper,
a vibrational energy
eigenstate, indexed by an integer $n$ (phonon number), was perfectly
correlated with the internal
states of a number of $N$ trapped ions. A read-out of the internal state
of
each ion would
yield a binary string, of length $N$ encoding a vibrational quantum
state
and this binary string
would occur with the same probability as the vibrational quantum number
in
the original
centre-of-mass state of the ion.  In this article  we  examine what
happens
if we readout only a
subset of the internal states of the trapped ions. We show that when the
vibrational mode is
prepared in a coherent state, then various  measurements on the ions
gives
various (in general
unequally  weighted) superpositions of  coherent states of different
phases.
In the limit where all
ions are measured the ring of superposed  coherent states results in a
Fock
state
\cite{janszky}. We show that this can actually be achieved with a {\em
single ion}, rather
than $N$ ions, if feedback is allowed in the computation. 
Detailed investigation of the phonon number measurement in terms
of  filter functions reveals a close relation to the 
Walsh function generators recently introduced in the context of 
searching a quantum database \cite{terhal}. We devote one section to 
exploring this relation.

%%%%%%%%%%%%%%%%%%%%%%%%%%%%%%%%%%%%%%%%%%%%%%%%%%%%%%%%%%%%%%%%%%%%%%%%%%%%

\section{The model}
\label{model}
In this section we summarize the model of D'Helon and Milburn 
\cite{dhelon}. The physical system consists of $N$ two-level 
ions forming a `crystallized' 
string confined in a linear rf trap. This is our quantum register. 
We make the usual assumptions 
that spontaneous emission from the upper internal level 
of the ions is negligible and that 
the ions to be cooled into the Lamb-Dicke limit. 
We also neglect decoherence of the center-of-mass motion. 
The electronic states of the ions are denoted by an integer $k$
%%%%%%%%%
\begin{equation}
|k\rangle_e = |S_N\rangle_N \otimes |S_{N-1}\rangle_{N-1} \otimes ...
\otimes |S_1\rangle_1 \,\, , 
\end{equation}
%%%%%%%%%%
where $S_i = 0, 1$ and
%%%%%%%%%%
\begin{equation}
k = S_N \times 2^{N-1} + S_{N-1} \times 2^{N-2} + ...+ S_1 \times 2^0.
\label{k}
\end{equation}
%%%%%%%%%%
The binary expansion for $k$ is thus just the string $S_N
S_{N-1}...S_1$. 

In the model of D'Helon et al.\cite{dhelon}, the interaction between the
centre-of-mass motion and
the internal electronic state of each ion was mediated by a detuned
(classical) laser pulse. 
For interaction times much greater than the vibrational period of the 
trap, the effective Hamiltonian for the $j$th ion is 
\begin{equation}
H_I^{(j)} = \hbar a^{\dagger}a  \chi
(\sigma_z^{(j)}+\frac{1}{2})
\hspace{2.5 mm},
\label{ham_far}
\end{equation}
where $\sigma_{z}^{(j)}$ is the population inversion for the $j$th 
ion and where $\chi=\eta_{CM}^2 
\Omega^2/(N\Delta)$. Here $\eta_{CM}$ is the Lamb-Dicke parameter, 
$\Omega$ is the Rabi 
frequency for the transitions between the two internal 
states of the ions,   
and $\Delta$ the detuning between the exciting laser
pulses and the
electronic transition. For simplicity we assume that these parameters 
are the same for all ions.

If we choose the durations $\tau_{j}$ of the 
standing wave pulses to increase geometrically with qubit number as
$\chi\tau_j=2^j\pi/2^N$, then the 
Hamiltonians $H_I^{(j)}$ generate a unitary transformation 
\begin{equation}
U=\exp\left(\frac{-i 2\pi a^{\dagger}a \Upsilon}{2^N}\right)
\hspace{2.5 mm},
\label{ut}
\end{equation}
where the electronic operator $\Upsilon$ 
provides a binary ordering of the qubits:
\begin{equation}
\Upsilon=\sum_{j=1}^N (\sigma_z^{(j)} + \mbox{$\frac{1}{2}$}) 2^{j-1}
\hspace{2.5 mm}.
\end{equation}
The eigenvectors of the operator $\Upsilon$ are the electronic number 
states
\begin{equation}
\Upsilon=\sum_{k=0}^{2^N-1} k |k \rangle_e \langle k|
\hspace{2.5 mm}.
\end{equation}

In the case of an $N$-ion register, the transformation $U$ does not
affect the 
vibrational
state, however it displaces the eigenstates of an electronic operator
$\Phi$ 
which is 
canonically conjugate to $\Upsilon$.
The eigenstates of $\Phi$ are a complementary set of electronic basis
states 
$\tilde{|p\rangle}_e$ given by Fourier transforms of the electronic
number 
states $|k\rangle_e$,
\begin{equation}
\tilde{|p\rangle}_e=\frac{1}{\sqrt{2}^N}\sum_{k=0}^{2^N-1} \exp(-2\pi i
k 
p 2^{-N}) |k\rangle_e
\hspace{2.5 mm},
\end{equation}
for $0 \leq p \leq 2^N-1$.  

We start with all the ions in the ground state $|0\rangle_e$ and 
an initially arbitrary vibrational state of all the ions, thus
%%%%%%%%%%%
\begin{equation}
|\psi_{in} \rangle = \sum_{n=0}^\infty c_n |n\rangle_{vib} \otimes
|0\rangle_e \,\, .
\end{equation}
%%%%%%%%%%%
We apply a Fourier transform ${\cal{F}}^-$ to the register, so that it
is 
in the state
%%%%%%%%%
\begin{equation}
{\cal{F}}^- {} |0 \rangle_e = | \tilde{0} \rangle_e 
\equiv \frac{1}{\sqrt{2^N}} \sum_{k=0}^{2^N -1} 
|k\rangle_e \,\, ,
\end{equation}
%%%%%%%%%
which is an equal superposition of all $|k\rangle_e$ states and 
the state of the system now reads
%%%%%%%%%
\begin{equation}
|\psi_{in}^\prime \rangle = \sum_{n=0}^\infty c_n |n\rangle_{vib}
\otimes 
|\tilde{0} \rangle_e \,\, .
\end{equation}
%%%%%%%%%
As explained in Ref.~\cite{dhelon} this can be easily achieved using 
$\pi/2$ laser pulses, one for each of the $N$ atoms.

We now apply the unitary transformation in Eq.(\ref{ut})
After this interaction we  arrive at an entangled state of the
vibrational 
mode and the register
%%%%%%%%%
\begin{equation}
|\psi_{out}^\prime \rangle = \sum_{n=0}^\infty \sum_{p=0}^{2^N -1}
c_{p+ n 2^N} |p + n 2^N\rangle_{vib} \otimes |\tilde{p}\rangle_e \,\, , 
\end{equation}
%%%%%%%%
where the register is in a superposition of Fourier transformed states. 
So the final step to get the desired result is now to apply the 
inverse Fourier transform ${\cal{F}}^+$ to the register. ${\cal{F}}^+$
satisfies ${\cal{F}}^+ {\cal{F}}^- = 1$. We get 
%%%%%%%%%%
\begin{equation} \label{entangle}
|\psi_{out}\rangle = {\cal{F}}^+ |\psi_{out}^\prime \rangle 
= \sum_{n=0}^\infty \sum_{k=0}^{2^N -1} c_{k+n2^N} |k +
n2^N\rangle_{vib}
\otimes |k\rangle_e \,\, , 
\end{equation}
%%%%%%%%%%
which is the basis for further investigations. 

%%%%%%%%%%%%%%%%%%%%%%%%%%%%%%%%%%%%%%%%%%%%%%%%%%%%%%%%%%%%%%%%%%%%%%%%%%
\section{Superpositions of coherent states on a circle}

In this section we assume that the vibrational state of the ions is 
a coherent state
%%%%%%%%%
\begin{equation}
|\psi \rangle _{vib} =  |\alpha\rangle_{vib} =
\sum_{n=0}^\infty c_n |n\rangle_{vib} = 
\sum_{n=0}^\infty \frac{\alpha^k}{\sqrt{k !}} e^{-\frac{1}{2}
|\alpha|^2} 
|n\rangle_{vib} \,\, . 
\end{equation}
%%%%%%%%%
Thus the final state after all the transformations described in section
\ref{model} is 
%%%%%%%%%
\begin{equation}
|\psi_{out} \rangle = \sum_{n=0}^\infty \sum_{k=0}^{2^N -1} 
\frac{\alpha^{k+n2^N}}{\sqrt{(k+n2^N) !}} 
e^{-\frac{1}{2} |\alpha|^2} |k + n2^N\rangle_{vib} \otimes |k\rangle_e
\,\, .
\end{equation}
%%%%%%%%%
What happens if we perform a measurement on say the first ion? 

We look at the definition of $k$, Eq. \ref{k}, and see that the first 
ion (i.e. ion number one) is determining whether $k$ is even or odd. 
So measuring ion number one is projecting the register in 
either a superposition of even or of odd Fock states, depending on the 
outcome of the measurement. Since the register is highly entangled 
with the vibrational state of the ions, this measurement will also 
have an effect on this state: It projects the former coherent state into 
either an even or odd Schr\"odinger cat state. This can be seen quite 
easily by writing the entangled state as a sum of states with even and 
odd $k$ 
%%%%%%%%%%
\begin{eqnarray}
|\psi_{out} \rangle & = & \sum_{n=0}^\infty \sum_{k=0}^{2^{N-1} -1}
\frac{\alpha^{2k+n2^N}}{\sqrt{(2k+n2^N) !}}
e^{-\frac{1}{2} |\alpha|^2} |2k + n2^N\rangle_{vib} \otimes |2k\rangle_e
\nonumber \\
 &  &  + {} \sum_{n=0}^\infty \sum_{k=0}^{2^{N-1} -1}
\frac{\alpha^{2k+1+n2^N}}{\sqrt{(2k+ 1+n2^N) !}}
e^{-\frac{1}{2} |\alpha|^2} |2k+1 + n2^N\rangle_{vib} \otimes 
|2k+1\rangle_e \,\, . 
\end{eqnarray}
%%%%%%%%%%
If we just want to get a Schr\"odinger cat state, we only need one ion, 
which means that $N=1$ and the even and odd cats are then entangled with 
the lower or upper internal level of this ion, respectively. Otherwise
we have to disentangle the states again, which can be done by 
applying ${\cal{F}}^+ \hat{U}^{-1} {\cal{F}}^-$ to the measured state. 

\begin{figure}[htb]
\centerline{\psfig{figure=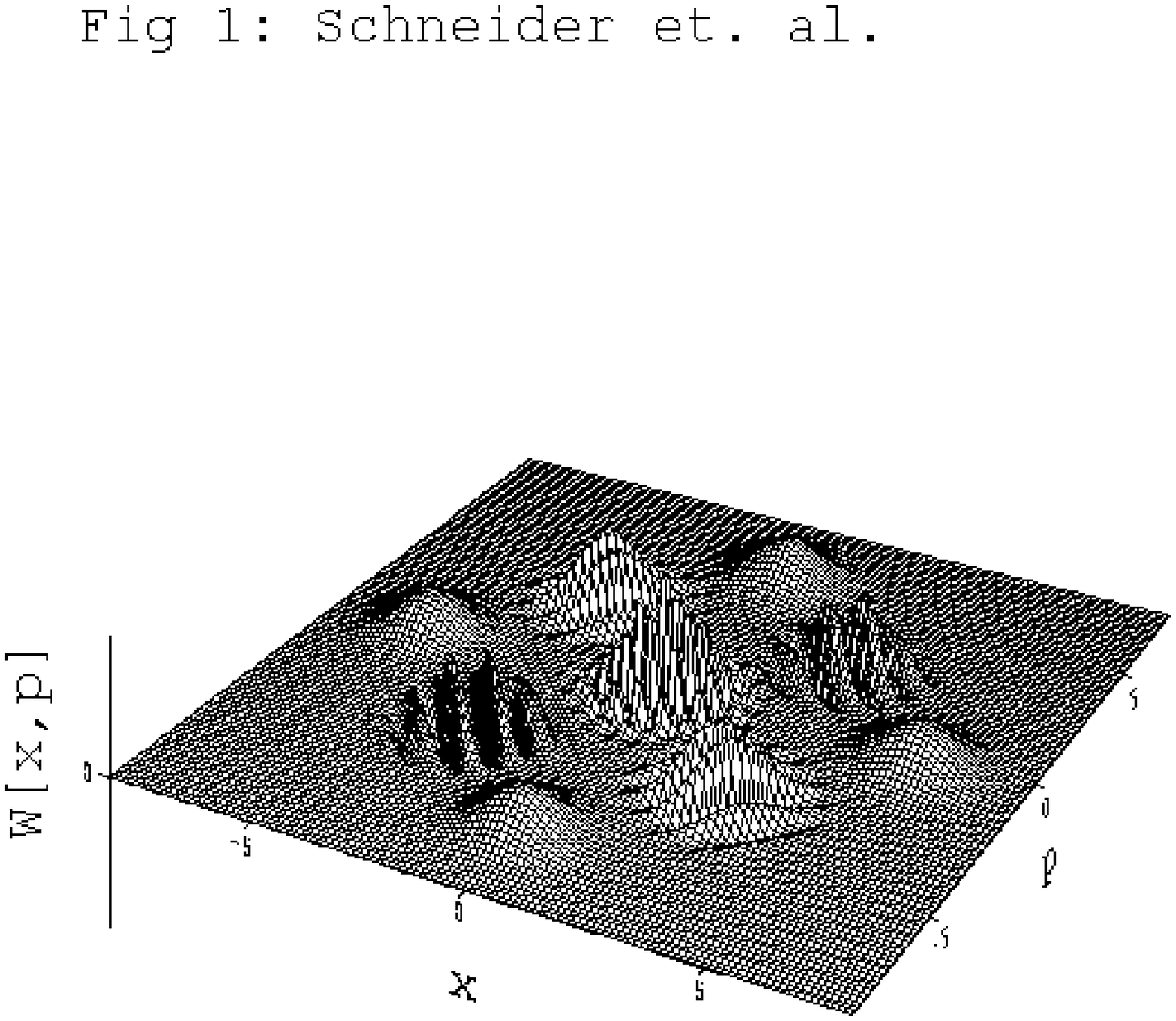,width=100mm}}
\caption{Wigner function for the state
$|\psi_{out}\rangle_{vib} \propto
|\alpha\rangle + |-\alpha\rangle + |i\alpha\rangle + |-i\alpha \rangle$
with $\alpha=4.0$.}
\label{fig1}
\end{figure}

So what happens if we do not stop after the  first ion but go on 
measuring the second, the third and so on? We just project into 
more and more superpositions of coherent states aligned around a circle 
of radius 
$|\alpha|$. 
Those superpositions are investigated in \cite{janszky}. They finally
lead
to a Fock state in the vibrational motion, depending on how many ions 
are used. For example, if the last two ions are measured and the 
result $S_{2}=S_{1}=0$ is obtained, we project the register in 
a superposition of $|4k\rangle_e$ states and the resulting vibrational
state is
%%%%%%%%%
\begin{equation}
|\psi_{out}\rangle_{vib} \propto |\alpha\rangle + |-\alpha\rangle + 
|i\alpha\rangle + |-i\alpha \rangle \,\, .
\label{state1}
\end{equation}
%%%%%%%%%%
The Wigner function for this is plotted in Fig.~\ref{fig1}.
Theses results are closely related to work done by Brune {\em et al.} 
\cite{brune90,brune92}. There the role of the ions is played by atoms 
and the phonons by photons in a cavity. 

It is interesting to note that the above measurement can in fact be done
with a 
{\em single} ion, providing we allow quantum-limited feedback in our 
computation. Instead of the simultaneous application of the unitary 
operator (\ref{ut}) on all ions (which requires $N$ different laser
pulses 
on the $N$ ions), one can apply $N$ different laser pulses on the same
ion. 
The scheme works as follows: The ion begins in the ground state
$|0\rangle$. 
We apply 
the Fourier transform operator to the ion, which is simply a $\pi/2$
pulse:
%%%%%%%%%%
\begin{equation}
{\cal F}^- |0\rangle_e  = |\tilde{0}\rangle_e  = (1/\sqrt{2}) 
( |0\rangle_e + |1\rangle_e ) \,\, .
\end{equation}
%%%%%%%%%%%
We now transform with the unitary operator 
%%%%%%%%%%
\begin{equation} \label{UU}
\hat{U}_j = \exp[-i2\pi \hat{a}^\dagger \hat{a} (\hat\sigma_z +1/2)
2^{-N+j-1}],
\end{equation}
%%%%%%%%%%
with $j=1$ initially. Then 
the $-\pi/2$ pulse is applied, producing the transformation ${\cal
F}^+$. 
Next, the atom state is read out in the $|0\rangle, |1\rangle $ basis.
The 
result is the least significant digit of $k$ (which determines whether
$k$ 
is even or odd). If the result is $1$ we apply the transform ${\cal
F}^+$, 
if $0$ we apply ${\cal F}^-$. It is these different unitary 
transformations, controlled by the result of a measurement, which 
constitute the feedback in the computation. 
We now move on to the next digit by 
doubling the interaction time of the pulse in Eq.~(\ref{UU}) by setting 
$j=2$. Following ${\cal F}^-$ we read out the next-to-least significant 
digit of $k$. The rest follows analogously until we reach $j=N$, where
$N$ in 
this case is completely arbitrary (there is only one ion!).

It might be thought that the measurement would be much quicker with 
simultaneous pulses on $N$ ions rather than consecutive pulses on a 
single ion, but this is in fact not the case, as can be seen as 
follows. Assuming that the $\pi/2$ pulses  can be achieved in 
negligible time, the time of the measurement will be dominated by the 
time taken to apply the unitary transformations (\ref{ut}) or
(\ref{UU}).
As pointed out above, to get the desired unitary transformation 
the interaction times have to be chosen so that 
%%%%%%%%%%%
\begin{equation}
\chi \tau_j = \frac{2^j \pi}{2^N} \,\, . 
\end{equation}
%%%%%%%%%%%
If the interactions are consecutive (whether on $1$ ion or $N$ ions), 
the total interaction time is 
%%%%%%%%%%
\begin{eqnarray}
t_{total} & = & \sum_{j=1}^N \tau_j =  \sum_{j=1}^N \frac{2^j \pi}{2^N
\chi}
= \sum_{j=1}^N \frac{2^j \pi N \Delta } {2^N (\eta_{CM} \Omega )^2}
\nonumber \\ &  = & 
 \frac{\pi N\Delta}{2^N (\eta_{CM} \Omega)^2} 2 \frac{1-2^N}
{1-2} = \frac{2 \pi N \Delta}{(\eta_{CM} \Omega)^2}(1 - 2^{-N}) 
\label{simN1}\,\, . 
\end{eqnarray}
If the interactions are simultaneous the total interaction time is 
just the longest interaction time
\begin{equation} \label{simN2}
t_{total} = \tau_{N} = \frac{ \pi N \Delta}{(\eta_{CM} \Omega)^2}
\end{equation}
which is at most a fraction of two smaller than the interaction time 
for consecutive pulses.

\begin{figure}[htb]
\centerline{\psfig{figure=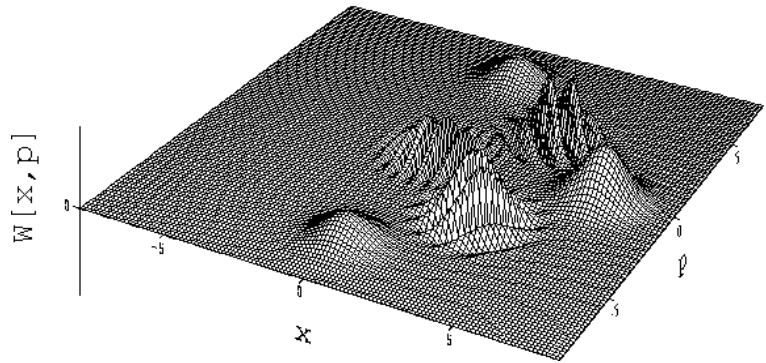,width=100mm}}
\caption{Wigner function for the state $|\psi_{out}\rangle_{vib} \propto 
2|\alpha\rangle +
(1-i)|i\alpha\rangle + (1+i)|-i\alpha\rangle$ with $\alpha=4.0$.}
\label{fig2}
\end{figure}

From this calculation we can conclude that 
for making the QND measurement of D'Helon and Milburn one can do 
perfectly well with one ion, which should make experimental
implementation much easier. However it is worth noting that there are
some experiments  which one could do with the 
entangled state (\ref{entangle}) which are not possible with a single 
ion. Assume that we do not measure 
the internal state of the first ion in the register, but perform our 
measurement instead on the second or third or any other ion, and then 
disentangle the remaining ions by applying ${\cal 
F}^{+}\hat{U}^{-1}{\cal F}^{-}$.
The result is not as simple  as 
an equally-weighted multiple-component cat-state, 
but has an interesting interference 
structure nevertheless. Essentially by 
measuring one ion (whichever) we set half of the $c_{k+n2^N}$ to 
zero.  For the first ion every second one is set to zero and the 
result is the standard cat state. 
For the second ion we get a pattern of $2^1$ states set to zero 
alternating with $2^1$ states not set to zero. In general measuring the 
$i^{\mbox{th}}$ ion creates a pattern consisting of $2^{i-1}$ states 
set to zero alternating with the same number of states not set to zero.
For 
example if we measure the second ion to be in the state $|0\rangle$, the 
register is in a superposition of states of the form $|4k \rangle_e$ 
and $|4k + 1\rangle_e$ ($k=0, 1, 2, ... 2^{N-2} -1$) and in this case 
we project the vibrational state into
%%%%%%%%%%
\begin{equation}
|\psi_{out}\rangle_{vib} \propto 2|\alpha\rangle + 
(1-i)|i\alpha\rangle + (1+i)|-i\alpha\rangle \,\, .
\label{state2}
\end{equation}
%%%%%%%%%%
The Wigner function for this state is plotted in Fig.~\ref{fig2}.

%%%%%%%%%%%%%%%%%%%%%%%%%%%%%%%%%%%%%%%%%%%%%%%%%%%%%%%%%%%%%%%%%%%%%%%%%%
\section{Searching for a Fock state}
 
The last considerations in the previous section have a close relation 
to classical searching as can be seen in \cite{terhal}. An unsorted
database
which has one marked entry (i.e.\ is of Hamming weight one) and contains 
$2^N$ items can 
be searched classically by bit strings of the same length but Hamming
weight
$2^{N-1}$. An answer of the database $y$ to the querie $g_k$, 
where $g_k$ is the hamming weight $2^{N-1}$ bit string which 
contains $2^k$ zeros alternation with $2^k$ ones, is a bit 
%%%%%%%%%%
\begin{equation}
a_k(g_k,y) = g_k \cdot  y = \left( \sum_{i=1}^n (g_k)_i y_i \right) 
\mbox{mod} \,\,\,   2 \,\, , 
\end{equation}
%%%%%%%%%
where $(g_k)_i$ and $y_i$ are the $i^{\mbox{th}}$ bits of $g_k$ and $y$ 
\cite{terhal}. 
Classically the answers of the database to $N$ queries with all the
possible 
Hamming weight $2^{N-1}$ bit strings uniquely determine the location 
of the marked entry. This location $l$ is just the number we get after 
transferring the ordered answers (i.e.\ a binary string of length $N$)
into their decimal form
%%%%%%%%%%
\begin{equation}
l = g_{N-1} \times 2^{N-1} + g_{N-2} \times 2^{N-2} 
+ ... + g_0 \times 2^0 \,\, .
\end{equation}
%%%%%%%%%
The bit strings $g_k$ play an essential rule in the recently 
introduced algorithm for searching a database \cite{terhal}. In this
work the 
authors search a database for one marked entry with a single querie
which 
consist of a superposition of Walsh-functions. 
Walsh-functions are all possible ``XOR'' combinations of the $g_k$s and 
the bit string containing $n$ zeros. Thus the $g_k$s are the generators 
of the Walsh-functions. 
Since in \cite{terhal} the answer to each Walsh-function is stored 
in the sign of the 
Walsh-function in the superposition state, further data processing 
to transform the information into $N$ bits is required. 

Now in our scheme we assume that the vibrational motion is in an 
unknown Fock state 
$|m\rangle_{vib}$ and our only knowledge about it is that 
$m \leq 2^N-1$. We choose to implement the measurement using $N$ ions  
in the trap. The state (\ref{entangle}) then reads
%%%%%%%%
\begin{equation}
|\psi_{out}\rangle = \sum_{k=0}^{2^N -1} c_k |k\rangle_{vib} \otimes 
|k\rangle_e \,\, , 
\end{equation}
%%%%%%%%%
where in this special case $c_k = \delta_{k,m}$ and so 
%%%%%%%%%%%%
\begin{eqnarray}
|\psi_{out}\rangle & = & |m \rangle_{vib} \otimes |m \rangle_e 
\nonumber \\ 
 & = & |m \rangle_{vib} \otimes |a_{N-1}(g_{N-1}, m)\rangle_{N} 
|a_{N-2}(g_{N-2}, m)\rangle_{N-1} ... |a_{0}(g_{0},m)\rangle_{1} \,\, . 
\end{eqnarray}
%%%%%%%%%%%%%
Measuring the register directly gives the Fock state we were looking for 
since each ion gives the answer of the database to a querie with 
a specific generator. This is different from other search
schemes \cite{terhal} where post processing is often required 
to obtain the information.

How fast can this measurement be done? The 
preparation of the superposition takes $O(\log(n))$ steps \cite{grover}. 
The interaction presented here also takes a time of order $O(\log(n))$. 
This is because both equation (\ref{simN1}) and equation 
(\ref{simN2}) imply a total time scaling as
\begin{equation}
t_{total} \sim N,
\end{equation} 
which is of order $\log(n)$. The inverse 
Fourier transform however takes $O((\log(n))^2)$ steps \cite{ekert}
which hence is the limiting process. So all 
in all our measurement scheme takes 
$O((\log(n))^2)$ steps. 

%%%%%%%%%%%%%%%%%%%%%%%%%%%%%%%%%%%%%%%%%%%%%%%%%%%%%%%%%%%%%%%%%%%%%%%%%
\section{Discussion}

In this paper we have investigated a measurement scheme for the
determination of the vibrational 
quantum number of the lowest normal mode of $N$ ions confined in a trap. 
The vibrational state of the ions becomes entangled with a state in the
product space of the
internal electronic states. This product state is defined by an integer
encoded as a binary string.
The measurement scheme
depends on the ability to perform a unitary Fourier transform operation on
the product space of the
internal electronic states of all the ions. This is a very large dimensional
unitary transformation
(of dimension $2^N$), but can be done if we regard the system as an ion trap
quantum computer. This
enables a significant generalisation of the standard von Neumann measurement
model. However we point
out that the same measurement can be accomplished with a single ion,
provided that subsequent
unitary transformations are selected on the basis of past measurements. This
is a kind of feedback
or adaptive scheme.

The measurement readout is done by reading out the internal electronic state
of each ion in some
predetermined order. If the excited state encodes a 1 and the ground state
encodes a 0, the result
of the measurement is a binary string which encodes an integer $n$. This result
will  occur with the same
probability as the occurrence of the normal mode vibrational quantum number
$n$ in the initial
centre-of-mass mode. The resulting conditional state of the centre-of-mass
is then a Fock state
$|n\rangle_{vib}$. If instead of reading out the entire binary string we
only readout a suitable
partition by making a state determination on a subset of ions, a number of
different vibrational
states can be prepared. We have shown for the case of an initial coherent
state of the normal mode
vibration that the resulting conditional states are in fact superpositions
of coherent states on a
ring of fixed radius in the phase-space of vibrational motion.

Determining selected partitions of a binary string is very similar to
methods used in binary
search routines. We have shown that our measurement scheme can in fact be
viewed as a kind of
quantum search algorithm for a particular integer $n$ encoding a vibrational
normal mode state.    

%%%%%%%%%%%%%%%%%%%%%%%%%%%%%%%%%%%%%%%%%%%%%%%%%%%%%%%%%%%%%%%%%%%%%%%%%%%
\section*{Acknowledgments}
S.~S. gratefully acknowledges financial 
support from a ``DAAD Doktorandenstipendium 
im Rahmen des gemeinsamen Hochschulsonderprogramms III von Bund und 
L\"andern'' and from the Center of Laser Science. 
%
%
%%%%%%%%%%%%%%%%%%%%%%%%%%%%%%%%%%%%%%%%%%%%%%%%%%%%%%%%%%%%%%%%%%%%%%%%%%%%

\newpage

\thebibliography{99}

\bibitem{ekert}
A.~Ekert and R.~Jozsa, Rev. Mod. Physics {\bf 68}, 733 (1996).

\bibitem{dhelon}
C.~D'Helon and G.~J.~Milburn, Phys. Rev. A {\bf 54}, 5141 (1996) and 
C.~D'Helon, \em Quantum-Limited Measurements on Trapped Laser-Cooled
Ions \em, PhD thesis, The University of Queensland, 1997.  

\bibitem{janszky}
J. Janszky \em et al. \em, Phys. Rev. A {\bf 48}, 2213 (1993).

\bibitem{terhal}
B.~M.~Terhal and J.~A.~Smolin, submitted to Phys. Rev. A, and 
Report No. quant-ph/9705041. 

\bibitem{brune90}
M.~Brune \em et al.\em, Phys. Rev. Lett. {\bf 65}, 976 (1990).

\bibitem{brune92}
M. Brune \em et al.\em, Phys. Rev. A {\bf 45}, 5193 (1992). 

\bibitem{grover}
L.~K.~Grover, Phys. Rev. Lett. {\bf 79}, 325 (1997).
\end{document}